\documentstyle[12pt,aaspp4,rotating]{article}
\def\msun{{\,M_\odot}}

\def\simlt{\lower.5ex\hbox{$\; \buildrel < \over \sim \;$}}
\def\simgt{\lower.5ex\hbox{$\; \buildrel > \over \sim \;$}}

\begin{document}
\centerline{To Appear in Astron. Nachr. Suppl. Issue 1/2003}
\vskip 0.5in
\title{A Relativistic Disk in Sagittarius A*}

\author{Siming Liu}
\affil{Center for 
Space Science and Astrophysics, Stanford University, Stanford, CA, 94305-4060}

\and

\author{Fulvio Melia}
\affil{Department of Physics and Steward Observatory, 
The University of Arizona, Tucson, AZ, 85721}
\altaffiltext{1}{NSF Graduate Fellow.}
\altaffiltext{2}{Sir Thomas Lyle Fellow and Miegunyah Fellow.}

\begin{abstract}

The detection of a mm/Sub-mm ``bump'' in Sgr A*'s radio spectrum suggests
that at least a portion of its overall emission is produced within a compact
accretion disk. This inference is strengthened by observations of strong
linear polarization (at the 10 percent level) within this bump. No linear
polarization has been detected yet at other wavelengths. Given that radiation
from this source is produced on progressively smaller spatial scales with
increasing frequency, the mm/Sub-mm bump apparently arises within a mere
handful of Schwarzschild radii of the black hole. We have found that 
a small (10-Schwarzschild-radii) magnetized accretion disk can not only
account for the spectral bump via thermal synchrotron processes, but that
it can also reproduce the corresponding polarimetric results. In addition, 
the quiescent X-ray emission appears to be associated with synchrotron 
self-Comptonization, while X-ray flares detected from Sgr A* may be 
induced by a sudden enhancement of accretion through this disk. The 
hardening of the flare-state X-ray spectrum appears to favor 
thermal bremsstrahlung as the dominant X-ray emission mechanism during 
the transient event. This picture predicts correlations among the mm, 
IR, and X-ray flux densities, that appear to be consistent with recent
multi-wavelength observations. Further evidence for such a disk in Sgr A*
is provided by its radio variability. Recent monitoring of Sgr A* at cm
and mm wavelengths suggests that a spectral break is manifested at 3 mm 
during cm/Sub-mm flares.

The flat cm spectrum, combined with a weak X-ray flux in the quiescent 
state, rules out models in which the radio emission is produced by 
thermal synchrotron process in a bounded plasma. One possibility is
that nonthermal particles may be produced when the large scale
quasi-spherical inflow circularizes and settles down into the small
accretion disk. Dissipation of kinetic energy associated with radial 
motion may lead to particle acceleration in shocks or via magnetic 
reconnection. On the other hand, the identification of a 106-day
cycle in Sgr A*'s radio variability may signal a precession of the 
disk around a spinning black hole. The disk's characteristics imply
rigid-body rotation, so the long precession period is indicative of
a small black-hole spin with a spin parameter $a/M$ around 0.1. It is
interesting to note that such a small value of $a/M$ would be favored 
if the nonthermal portion of Sgr A*'s spectrum is powered by a 
Blandford-Znajek type of process; in this situation, the observed 
luminosity would correspond to an outer disk radius of about 30 
Schwarzschild radii. This disk structure is consistent with earlier
hydrodynamical and recent MHD simulations and is implied by Sgr A*'s
mm/Sub-mm spectral and polarimetric characteristics.

For the disk to precess with such a long (106-day) period, the angular
momentum flux flowing through it must be sufficiently small that any
modulation of the total angular momentum is mostly due to its coupling 
with the black-hole spin. This requires that the torque exerted on the 
inner boundary of the disk via magnetic stresses is close to the angular 
momentum accretion rate associated with the infalling gas. Significant 
heating at the inner edge of the disk then leaves the gas marginally 
bounded near the black hole. A strong wind from the central region may 
ensue and produce a scaled down version of relativistic (possibly
magnetized) jets in AGNs.

\end{abstract}



\section{Introduction}

The compact radio source, Sgr A*, at the dynamical center of our Milky Way 
Galaxy, is believed to be associated with a supermassive black hole (Melia
\& Falcke 2001). Evidence in support of this is quite compelling, 
especially with the detection of a 3 hour X-ray flare (Baganoff et al. 2001) 
from the direction of Sgr A* and the recent monitoring of a star 
orbiting within light-days of the black hole, which points to a central
mass of $3.7\pm1.5 \times 10^6\msun$ (Sch\"{o}del et al 2002), consistent 
with an earlier measurements of $\sim 2.6\times 10^6\msun$ (Eckart \& 
Genzel 1996; Ghez et al. 1998).

The nature of Sgr A* thus bears critically on our understanding of black-hole
physics. Several mechanisms have been proposed over the past decade to explain
its broad band spectrum and polarization, among them Bondi-Hoyle accretion 
(Melia 1992); a two-temperature, viscous disk (Narayan, Yi \& Mahadevan 1995); a
relativistic nozzle (Falcke \& Markoff 2000) and a recent combination of
advection-dominated disk with a nozzle (Yuan, Markoff \& Falcke 2002). Over
the past decade or so, our group has adopted a theoretically-motivated 
phenomenological approach (Melia, Liu \& Coker 2000, 2001), in which the
observations play a crucial role in constraining the theoretical picture.

The detection and confirmation of a mm/Sub-mm bump in Sgr A*'s spectrum
suggests an emission component different from that responsible for
the cm radio emission (Zylka, Mezger \& Lesch 1992; Falcke et al. 1998).  
This emission component is also implied by Sgr A*'s variability.  
Radio observations show that Sgr A*'s fluctuation amplitude increases
toward high frequency (Zhao \& Goss 1993) and there is a spectral break at
3 mm during radio flares (Zhao et al. 2003). Since high-frequency radio
emission is produced by relatively more energetic particles, located deeper 
in the gravitational well of the black hole (Melia, Jokipii \& Narayanan 1992), 
the mm/Sub-mm emitting gas should be very close to the black hole's event horizon.

The detection of linear polarization in this spectral bump enhances this
inference further and sets severe constraints on possible explanations for 
this component (Aitken et al. 2000). No significant linear polarization has 
yet been detected at frequencies lower than $112$ GHz, though relatively
strong circular polarization persists in the cm band (Bower et al. 2002). 
The flip of the position angle of the polarization vector by
about $90^\circ$ between $230$ GHz and $350$ GHz favors a scenario where 
the mm/Sub-mm emission is produced within a small, optically thin, 
magnetized accretion disk (Melia et al. 2000). No other model so far can 
explain this linear polarization characteristic (In the empirical
model of Agol (2000), the frequency where the position angle flips by 
$90^\circ$ is much lower than the frequency corresponding to the 
spectral peak of the flux density, which is not in line with the observations).

The existence of a small disk is also motivated theoretically. Earlier 
hydrodynamical simulations suggested that black-hole accretion from stellar 
winds, as is the case for Sgr A*, is characterized by a small angular 
momentum of the captured gas (Coker \& Melia 1997). This accreted angular 
momentum is too small for the gas to settle onto a large disk, as required 
by the ADAF model (Narayan et al. 1995). The captured angular momentum is
instead barely sufficient to circularize the gas just before it falls 
across the black hole's event horizon. Detailed MHD simulations have
provided an indication of the structure for such a disk (Hawley \& Balbus 
2002). The fact that the Magneto-Rotational Instability (Balbus \& Hawley 
1991) can induce a MHD dynamo in the disk provides a straightforward
explanation for the mm/Sub-mm bump as the result of synchrotron process. 
The magnetic field also provides an anomalous viscosity and, given its
strength, couples the electrons and ions via reconnection, so that
a single temperature fluid is maintained (Melia et al. 2001).

X-ray observations of Sgr A* have provided additional means of learning
about its nature.  It turns out that Sgr A* is an extremely weak X-ray 
source with a quiescent X-ray luminosity of $2.2^{+0.4}_{-0.3}\times 10^{33}$ 
erg s$^{-1}$ (Baganoff et al. 2001). Interestingly, electrons responsible 
for the mm/Sub-mm emission also Comptonize the radio photons into the
X-ray band. It is notable that the physical conditions required to produce 
the mm/Sub-mm spectrum can also account for Sgr A*'s quiescent X-ray emission. 

{\it Chandra} also detected a strong X-ray flare from the direction of Sgr 
A*. The flare lasted about 3 hours and featured a variation on a 10 minute 
time scale, suggesting an emission region no bigger than $20\,r_S$, where 
$r_S= 7.7\times 10^{11}$ cm is the Schwarzschild radius for the $2.6\times 
10^6\msun$ supermassive black hole associated with Sgr A*. The peak flux 
density for this flare is 50 times higher than that in the quiescent state. 
Recent X-ray observations have shown that this type of X-ray flare is 
common to Sgr A*, occurring about once per day (Baganoff et al. 2003; 
Goldwurm et al. 2003). 

The fact that the flares are very strong, are variable on a 10 minute time
scale and have a flat spectrum, pose telling theoretical challenges which, 
at the same time, also create a valuable opportunity for constraining the
physical conditions near the black hole's event horizon. Our study has
shown that an enhancement of the mass accretion rate through the disk can 
not only account for these flares, but can also induce strong Sub-mm/Far-IR 
flares that should occur simultaneously with the X-ray flares (Liu \& Melia 
2002a). Recent observations of Sgr A*'s mm/Sub-mm variability has indicated 
that the Sub-mm spectral index increases significantly during radio flares 
(Zhao et al. 2003), consistent with our prediction.  However, radio flares 
usually last for several days, which is much longer than the duration of 
an X-ray flare, suggesting more complicated physical processes. Nevertheless, 
the nondetection of an IR flare (Hornstein et al. 2002) seems to favor this 
model, where X-ray flares are produced via thermal bremsstrahlung processes, 
over the nozzle model, in which synchrotron self-Comptonization is introduced 
to account for the X-ray flare emission (Markoff et al. 2001).

Moreover, the low quiescent X-ray flux also delimits hot gas content 
around Sgr A*. When this constraint is combined with the flat radio spectrum,
one can show that the cm radio emission from Sgr A* cannot be produced by a
bounded, thermal synchrotron source (Liu \& Melia 2001).  One possibility 
is that the radio emission is produced via nonthermal synchrotron processes 
in the region where the large scale quasi-spherical inflow circularizes to 
form the small accretion disk responsible for the mm/Sub-mm and X-ray 
emission. Energetic, nonthermal electrons can in principle be produced
by the dissipation of kinetic energy associated with the radial motion 
of the infalling gas in shocks or magnetic reconnection. Assuming that 
a fixed fraction of particles is accelerated in this way, one can
obtain a good fit to the radio spectrum. The circular polarization 
properties may then be associated with the turbulent nature of the gas 
in this region (see, e.g., Beckert \& Falcke 2002; Ruszkowski \& Begelman 
2002).

On the other hand, a 106 day period in Sgr A* radio variability recently
reported by Zhao et al. (2001) appears to be associated with the precession
of a small hot disk under the influence of a spinning black hole (Liu \&
Melia 2002b). The physical characteristics of the disk indicate that it
will precess as a rigid-body. However, for the disk to survive longer than
the observed period, the net angular momentum flux through the disk must
be extremely small, which requires that the inward angular momentum flux
associated with the accreting gas must be cancelled almost completely by
the outward angular momentum induced by torque associated with the
magnetic stresses. A nonzero torque at the inner edge of an accretion disk
has been discussed extensively (Krolik 1999; Gammie 1999; Agol \& Krolik
2000) during the past few years. Recent MHD simulations have also
confirmed several of these theoretical speculations (Hawley \& Balbus 2002).
Should this picture be correct, it should be noted that a small black hole 
spin of $\sim 0.1$ $M$, where $M$ is the mass of the black hole, would be 
favored if the nonthermal portion of Sgr A*'s spectrum is instead powered 
by energy extracted from the black hole via a Blandford-Znajek type of
process. The precession period then requires that the disk has an outer 
radius of $\sim 30 r_S$, consistent with the general picture described 
above. The power extracted from the black hole also heats up the gas near 
the event horizon and unbinds it. The ensuing wind is not unlike the
relativistic jets observed in AGNs. Further exploration of this idea
may eventually reveal a more refined view of the processes hidden in
the central engine of these sources.

\section{A Relativistic Disk Model for the mm/Sub-mm Emission from Sgr A*}

\begin{figure}[htb]
\begin{minipage}[t]{.45\textwidth}
\includegraphics[width=\textwidth, height=6cm]{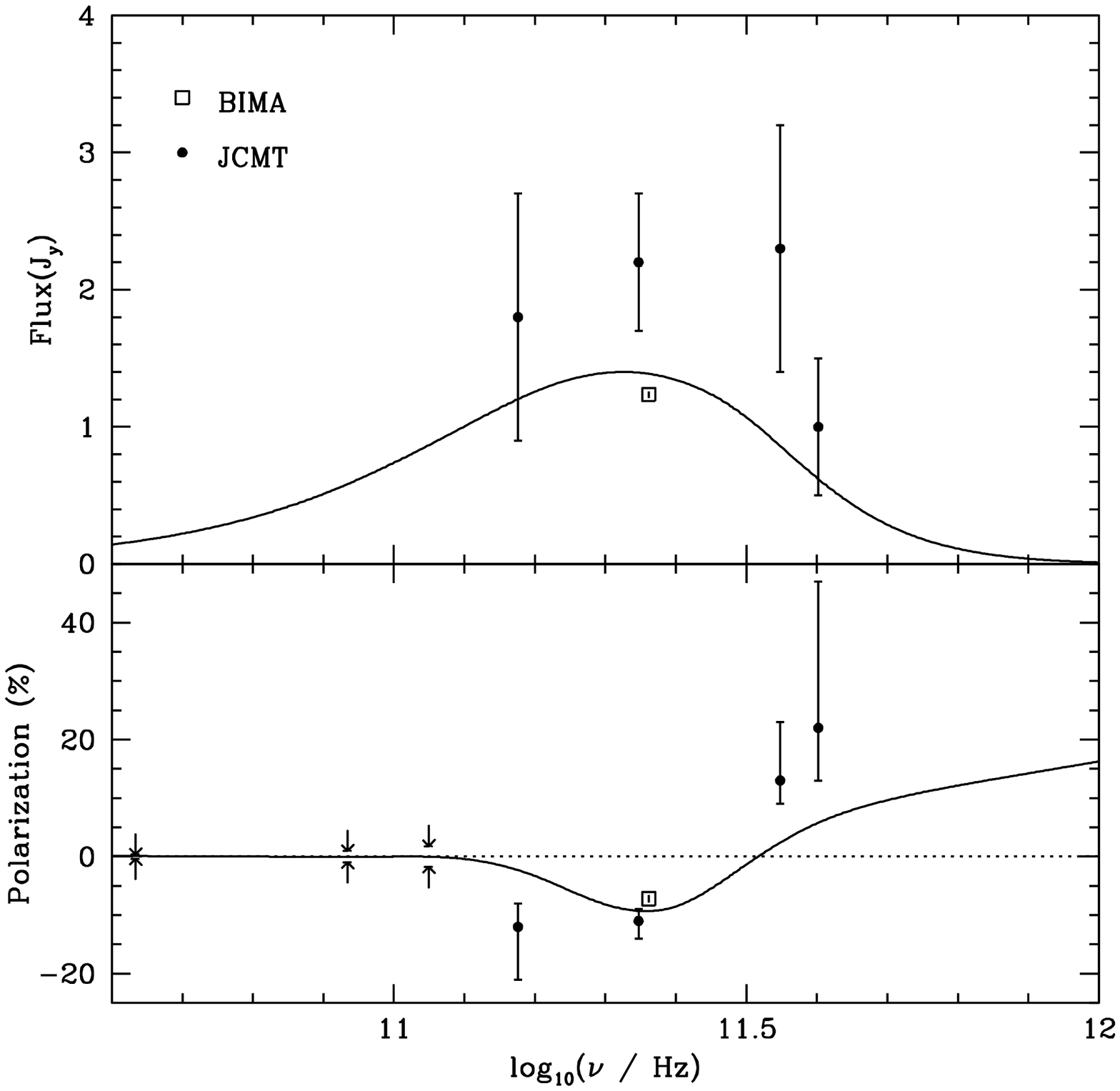}
\caption{
Best fit to the linear polarization of radio emission from Sgr A*. Here 
the model parameters are as follows: $\dot{M} = 4.1\times 10^{16}$g 
s$^{-1}$, $\beta_p=0.02$, $\beta_\nu=0.2$, $r_i = 1.8\, r_S$ and 
$r_o=8.5\,r_S$. The inclination angle of the disk is $30^\circ$. At the 
outer boundary, the gas temperature is fixed by the assumption that the
thermal energy of the gas equals $7\%$ of its dissipated gravitational 
energy.} 
\label{SimingL-polar.ps} 
\end{minipage} 
\hfil
\begin{minipage}[t]{.45\textwidth}
\includegraphics[width=\textwidth, height=5.7cm]{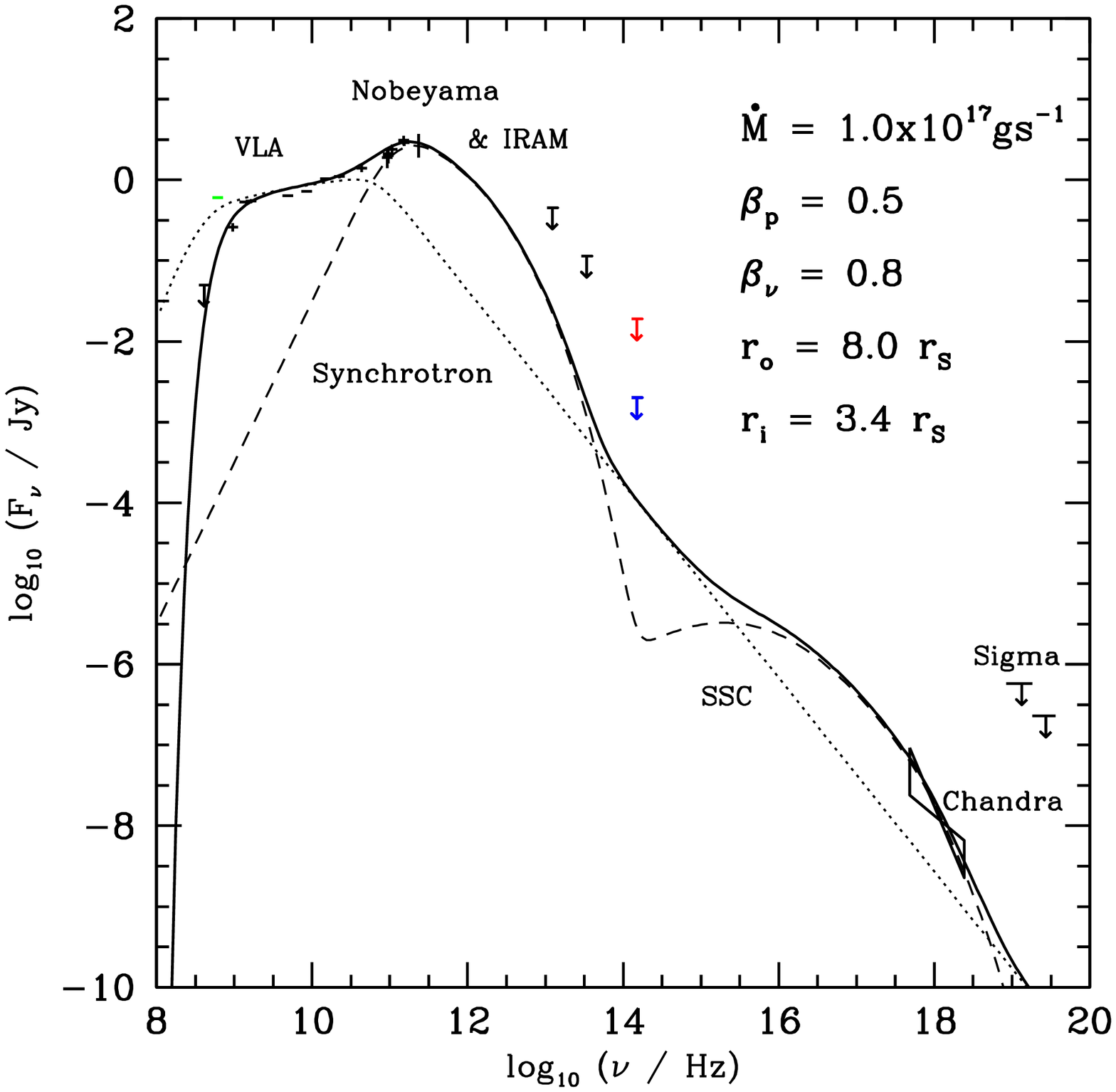}
\caption{
Best fit to Sgr A*'s quiescent spectrum. The model parameters are 
shown in the figure. Here the disk has an inclination angle of $45^\circ$.
The thermal energy of the gas is assumed to equal $80\%$ of its
dissipated gravitational energy at $r_o$. The dashed line here denotes 
emission from the small disk. The dotted line gives emission produced by 
nonthermal particles in the circularization zone.
}
\label{SimingL-radio.ps}
\end{minipage}
\end{figure}

The model of a hot, magnetized, small accretion disk in Sgr A* has been 
developed fully in the paper by Melia et al. (2001) where, prior to the
availability of all the observational constraints described above, the 
inner boundary condition was chosen to have zero torque. In this instance,
the temperature at the outer boundary of the Keplerian region is the
primary free parameter. The disk structure is determined once one specifies 
the inner ($r_i$) and outer ($r_o$) radii, the magnetic ($\beta_p$) and viscous 
($\beta_\nu$) parameters, the mass accretion rate $\dot{M}$ and the inclination 
angle of the disk. The best fit to the mm/Sub-mm polarization and spectral
data is shown in Figure \ref{SimingL-polar.ps} (Melia et al. 2000). Note that
here a negative percentage means that the position angle of the 
polarization vector is parallel to the angular momentum vector of the 
disk, while positive polarization means that the polarization vector flips 
by $90^\circ$ with respect to negative polarization. The frequency at which
the polarization vector flips is $\sim 3.3\times 10^{11}$ Hz, which is 
higher than the peak frequency of the flux density, $\sim 2.1\times 
10^{11}$ Hz. This is a unique feature of our relativistic disk model, 
that is apparently not yet matched by alternative scenarios (cf. 
Agol 2002).

It is straightforward to understand these polarization characteristics. At 
mm wavelengths, the red shift side of the disk becomes optically thin first. 
At this point, the emission is mostly from the front and back of the black 
hole, where it is polarized in the direction parallel to the disk's spin 
axis due to the influence of the very strong toroidal field within the 
disk. At Sub-mm frequencies, even the gas to the front and back of the 
black hole becomes optically thin, and the emission from the blue shifted 
side of the disk dominates; the polarization vector thus flips by $90^\circ$. 
Faraday rotation by the intervening plasma will make the observed flip of 
the polarization vector different from $90^\circ$, which can reconcile the 
slight difference between the theoretical prediction and the observational 
results. Due to the relatively poor angular resolution of JCMT ($22"$ at 220
GHz), the corresponding error bars are quite big, as can be seen from
Figure \ref{SimingL-polar.ps} (Aitken et al. 2000). However, the detection of
strong linear polarization is quite obvious. Recent high resolution
($3.6"\times 0.9"$) BIMA observations have confirmed strong linear
polarization at $220$ GHz (Bower et al. 2003), adding some confidence
to the model.

\section{X-ray Emission from the Relativistic Disk}

{\it Chandra} observations indicate that quiescent X-ray emission from Sgr
A* is very weak and soft, with an X-ray luminosity $2.2^{+0.4}_{-0.3}\times 
10^{33}$ erg s$^{-1}$ and a spectral index $1.5^{+0.8}_{-0.7}$, which is 
not consistent with an ADAF (Baganoff et al.  2001). Given the fact that 
the plasma is so hot in the disk that electrons are relativistic (Melia 
et al. 2001), one is motivated to investigate the effects of synchrotron 
self-Comptonization (SSC) in this medium.
Radio variability observations and recent theoretical developments 
(e.g., Liu \& Melia 2002b) favor a nonzero stress at the inner edge of the 
accretion disk. We will henceforth adopt a zero angular momentum flux condition.
The radial velocity of the accretion flow is then $v_r = -\beta_\nu \beta_p 
R_g T/\mu r \Omega\,$.  The other equations derived in Melia et al. (2001) 
are still applicable and we won't reproduce them here. In Figure \ref{SimingL-radio.ps}, 
we provide the best fit to Sgr A*'s broadband spectrum. SSC evidently accounts
for Sgr A*'s quiescent X-ray spectrum very well.

\begin{figure}[!htb]
\includegraphics[width=1.0\textwidth, height=8cm]{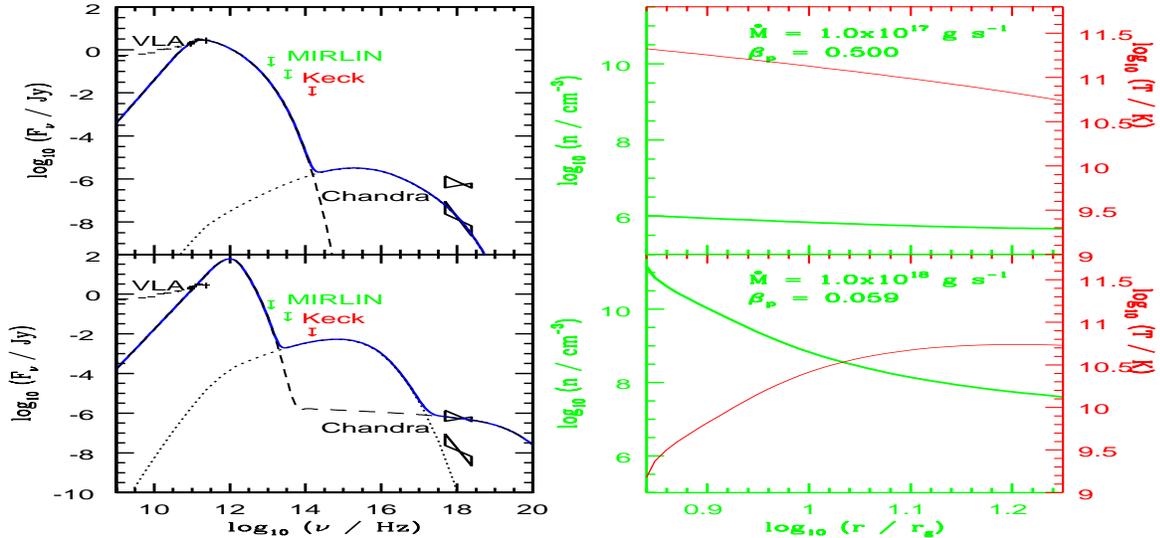}
\caption{
An accretion induced X-ray flare from Sgr A*. The righthand panels give the 
temperature (thin lines) and density profiles (thick lines) for the disk ($r_g=r_S/2$). 
The left panels show the corresponding disk spectra.  Note that the model 
prediction is consistent with the IR upper limits (Hornstein et al. 2002).
} \label{SimingL-flare.ps}
\end{figure}

However, Sgr A*'s X-ray emission during a flare is much more complicated
than that during quiescence. Although the short variation time scale of 
10 minutes is consistent with the flare being induced by an accretion process
in the disk, the fact that the X-ray flux density can increase by a factor 
of $50$ suggests dramatic changes in the disk's structure. Moreover, the 
hardening of the X-ray spectral index also rules out SSC as the dominant 
X-ray emitting mechanism during the flare (Liu \& Melia 2002a). We note 
that the hot disk described here is not stable when the mass accretion 
rate is large. When $\dot M$ increases, bremsstrahlung cooling becomes 
more and more important and can be the dominant cooling mechanism. The 
X-ray flare may in fact be associated with enhanced thermal bremsstrahlung 
emission.

Figure \ref{SimingL-flare.ps} depicts such a scenario. Here the disk has an outer
radius of $9\ r_S$ and we assume that the enhancement of accretion through it
can suppress the MHD dynamo. A justification for this is that cooling 
becomes more efficient with increasing $\dot M$, and this decreases the gas 
temperature. We infer that $\beta_p$ is thus anti-correlated with $\dot{M}$. 
During the X-ray flare, the gas density can be as high as $10^{10}$ cm$^{-3}$ 
and the magnetic field reaches $100$ Gauss. The corresponding synchrotron 
cooling time is a few hours, consistent with the general picture outlined 
above (Petrosian 1985).  Recent multi-wavelength observations have indicated 
that there is no obvious flux change at 3 mm during the X-ray flare (Baganoff et al. 2003). 
According to our model, the disk is optically thick at 3 mm, so the flux 
density is not expected to change significantly (see Figure 3).

\section{The Nature of Radio Emission from Sgr A*}

In Figure \ref{SimingL-radio.ps} we showed a fit to the cm emission from Sgr A* 
under the assumption that this radiation is produced via nonthermal 
synchrotron processes in the circularization zone (the model details 
can be found in Liu \& Melia 2001). The distribution of nonthermal 
particles is given by $N(E,r) = 1.7\times 10^{-11} E^{-3.4} 
n(r),$ where $E$ is the electron energy, and $n(r)$ is the electron density 
at radius $r$. Although magnetic reconnection at smaller radii of the 
disk can also induce particle acceleration (adding to the contribution
made by the nonthermal particles in the circularization zone), the fact 
that the gas temperature is as high as $10$ Mev there appears to make the 
thermal process dominant. Nevertheless, a complete treatment of this 
problem incorporating particle acceleration via magnetic reconnection is 
warranted. 

We can understand the nonthermal nature of cm radio emission using the 
following argument. Because quiescent X-ray emission from Sgr A* is 
weak (see Figure \ref{SimingL-radio.ps}), we can constrain the hot gas content 
in Sgr A* via its bremsstrahlung emissivity. If we assume that the gas 
is bounded, the gas temperature must be lower than its virial value. 
Combining these two upper limits, one can show that to produce the $1.36$ 
GHz flux from Sgr A* the magnetic field energy density must be more than 
ten times bigger than the thermal energy density of the hot gas. Such a 
configuration is not physical if the magnetic field is intrinsic to the 
hot gas. Of course, it is also possible that the radio emission is 
produced by some unbounded plasma, as proposed in the jet model by Markoff 
and Falcke (2000). Then the origin of the jet becomes the large unknown
in the model.

\begin{figure}[htb]
\begin{minipage}[t]{.45\textwidth}
{
\vspace{-8.3cm}

\hspace{0.4cm}The detection of a 106 day radio cycle is intriguing 
because it is intrinsic to Sgr A* (Zhao et al. 2001) and recent VLA 
observations indicate that the emission is produced within $140 r_S$ 
(Bower et al. 2002). The dynamical time scale within such a small region 
is much shorter than this period, suggesting it may 
be associated with an intrinsic property of the black hole. Here we discuss 
the possibility of accounting for the periodicity with precession of 
the disk around a spinning black hole. Because the disk is hot and 
magnetized, the strong internal coupling will invalidate the 
Bardeen-Petterson effect (1975). However, to make the disk survive 
over a time scale longer than the observed period, the angular momentum 
flux through the disk must be close to zero (Liu \& Melia 2002b). Under
these conditions, it is straightforward to calculate the precession 
period $P$ of such a disk around a spinning black hole. 
}
\end{minipage}
\hfil
\begin{minipage}[t]{.45\textwidth}
\includegraphics[width=\textwidth]{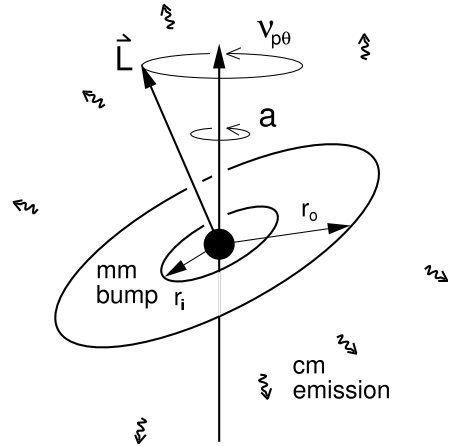}
\vspace{-3cm}
\caption{
Schematic diagram of a precessing compact disk around the
supermassive black hole in Sgr A*. Mm/Sub-mm to X-ray emission from Sgr 
A* are mostly produced within the disk. The cm radio emission, on the 
other hand, may be produced by diffusive nonthermal particles energized 
near the black hole.}
\label{SimingL-precess.ps}
\end{minipage}
\end{figure}
\vspace{-0.4cm}
Considering the 
fact that the disk surface is roughly a constant, we have
$P = \pi r_o^{5/2}r_i^{1/2}[1-(r_i/r_o)^{5/2}]/ \\
5aM [1-(r_i/r_o)^{1/2}]\,.$ Setting P=106 days, we have
$a/M = 0.088 (r_i/3r_S)^{1/2} (r_o/30r_S)^{5/2} \{0.69[1-(r_i/r_o)^{5/2}]/ 
[1-(r_i/r_o)^{1/2}]\}\,.$
The precession of a larger structure would require a bigger black hole 
spin. We note that power extracted via a Blandford-Znajek type of process 
(Blandford \& Znajek 1977) is then given by:
$L\sim1.1\times 10^{34}(B/29{\rm Gauss})^2(a/0.088M)^2(M/2.6\times 
10^6\msun)^2 {\rm erg\, s}^{-1}\,,$
which is very close to Sgr A*'s radio luminosity. If this extracted power 
is responsible for radio emission from Sgr A*, then the structure of 
Sgr A* should be like that depicted in Figure \ref{SimingL-precess.ps}. 

We thus have the following picture for Sgr A*. Stellar winds in and around 
the Galactic center may be captured by the supermassive black hole. The 
small angular momentum of the captured gas leads to the formation of a 
small hot accretion disk near the event horizon. Mm/Sub-mm to X-ray emission 
is mostly produced within the small disk. Before the hot plasma falls into 
the black hole, however, a certain fraction of the gas is accelerated (nonthermally), 
either via energy extracted from a spinning black hole, or by energy liberated
below the marginal stable orbit. These energetic particles then diffuse to 
larger radii.  The flat radio spectrum is produced by these particles. The 
precession of the disk around the spin of the black hole can lead to a corresponding
modulation of the outflow perceived in projection as well, inducing a periodic 
variation of the radio flux density.

\section{Conclusions}

Our study has shown that a hot, magnetized, relativistic accretion disk plays 
an essential role in revealing the nature of Sgr A*. Future observations of 
this source at $690$ GHz (private communication with J. H. Zhao), combined 
with current X-ray observations, will help us to understand the interaction 
between the disk and the black hole,  which is believed to be a key element 
of all AGNs. A comprehensive investigation of the supermassive black hole 
at the Galactic Center may eventually help us to unravel the mysterious inner
workings of the most powerful engines at the nuclei of active galaxies.

\section{acknowledgement}

S. Liu thanks the organizing committee of the 2002 GC conference for 
financial assistance. This research was supported by NASA grants NAG5-8239 
and NAG5-9205 at Arizona, and by the Center for Space Science and 
Astrophysics at Stanford.



\begin{thebibliography}{40}

\bibitem{Agol00} Agol, E. 2000, ApJ, 538, 121L

\bibitem{AgolK00}Agol, E., \& Krolik, J. 2000, ApJ, 528, 161

\bibitem{Aitken00}Aitken, D.K. et al., 2000, ApJ, 534, L173
 
\bibitem{Baganoff01}Baganoff, F. et al., 2001, Nature, 413, 45

\bibitem{Baganoff03} Baganoff, F. et al. these proceedings eds. Falcke, H. 
et al. 2003

\bibitem{BH91} Balbus, S.A., \& Hawley, J.F. 1991, ApJ, 376, 214

\bibitem{Bardeen75}Bardeen, J.M., \& Petterson, J.A. 1975, ApJ, 195, 
L65

\bibitem{Beckert02}Beckert, T., \& Falcke, H. 2002, A\&A, 388, 1106

\bibitem{Blandford77}Blandford, R., \& Znajek, R. 1977, MNRAS, 179, 433

\bibitem{Bower02}Bower, G., Falcke, H., Sault, R.J., \& Backer, D.C.
2002, ApJ, 571, 843

\bibitem{Bower03}Bower, G., Falcke, H., Sault, R.J., \& Backer, 
D.C. 2003, ApJ, 588, 331

\bibitem{Coker97}Coker, R.F., \& Melia, F. 1997, ApJ, 488, L49

\bibitem{Eckart96} Eckart, A., \& Genzel, R. 1996, Nature, 383, 415

\bibitem{FM00} Falcke, H., \& Markoff, S. 2000, A\&A, 362, 113

\bibitem{Falcke98}Falcke, H., Goss, W.M., Matsuo, H., Teuben, P.,
Zhao, J.H., \& Zylka, R. 1998, ApJ, 499, 731

\bibitem{Gammie99}Gammie, C. 1999, ApJ, 522, L57

\bibitem{Ghez98} Ghez, A.M. et al., 1998, ApJ, 509, 678

\bibitem{Goldwurm02}Goldwurm, A. et al., 2003, ApJ, 584, 751

\bibitem{HB02}Hawley, J.F., \& Balbus, S.A. 2002, ApJ,573, 738

\bibitem{Horn02}Hornstein, S.D. et al., 2002, ApJ, 577, L9

\bibitem{Krolik99}Krolik, J. 1999, ApJ, 515, L73

\bibitem{Liu01}Liu, S., \& Melia, F. 2001, ApJ, 561, L77
 
\bibitem{Liu02a}Liu, S., \& Melia, F. 2002a, ApJ, 566, L77

\bibitem{Liu02b}Liu, S., \& Melia, F. 2002b, ApJ, 573, L23

\bibitem{Melia92}Melia, F. 1992, ApJ, 387, L25

\bibitem{MeliaFalcke01}Melia, F., \& Falcke, H. 2001, ARA\&A 39, 309

\bibitem{Meliaetal92}Melia, F., Jokipii, R., \& Narayanan, A. 1992, 
ApJ, 395, L87

\bibitem{Melia00}Melia, F., Liu, S., \& Coker, R. 2000, ApJ, 545, L117
 
\bibitem{Melia01} Melia, F., Liu, S., \& Coker, R. 2001, ApJ, 553, 
146

\bibitem{Narayan95}Narayan, R., Yi, I., \& Mahadevan, R. 1995, Nature, 
374, 623

\bibitem{Petrosian85}Petrosian, V. 1985, ApJ, 299, 987

\bibitem{Ruszkowski02} Ruszkowski, M., \& Begelman, M.C. 2002, ApJ, 
581, 223

\bibitem{Schodel02}Sch\"{o}del, R. et al., 2002, Nature, 419, 694

\bibitem{Yuan02} Yuan, F., Markoff, S., \& Falcke, H. 2002, A\&A, 
383, 854

\bibitem{Zhao93}Zhao, J.H., \& Goss, W.M. in: Sub-arcsecond radio 
astronomy (Cambridge: Cambridge Univ. Press, 1993), p.38

\bibitem{Zhao01}Zhao, J.H., Bower, G.C., \& Goss, W.M. 2001, ApJ,
547, L29
 
\bibitem{Zhao03} Zhao, J.H. et al., 2003 ApJ, 586, 29

\bibitem{Zylka92} Zylka, R., Mezger, P., \& Lesch, H. 1992, A\&A, 
261, 119

\end{thebibliography}
\end{document}